\documentclass[
aps,%
12pt,%
final,%
notitlepage,%
oneside,%
onecolumn,%
nobibnotes,%
nofootinbib,%
superscriptaddress,%
noshowpacs,%
centertags]%
{revtex4}

\newcommand{\isn}[2]{\mbox{$^{#2}${#1}}}
\newcommand{\WI}[2]{#1_{\mathrm{#2}}}

\begin{document}

\title{Light Neutral Clusters in Supernova Matter}

\author{{\bf \hspace{-1.3cm} 2019} \firstname{I.V.}~\surname{Panov$^{1),2) *}}$,
}

\footnotetext[1]
{\it Institute for Theoretical and Experimental Physics, National
Research Center Kurchatov Institute, Moscow, 117218
Russia. }
 \footnotetext[2]{\it
 National Research Center Kurchatov Institute, Moscow,
123182 Russia \\ \rm ${^{* }}$ Igor.Panov@itep.ru\\  ${^{** }}$ Yudin@itep.ru
}%
\author{\firstname{A.V.}~\surname{Yudin$^{1) **}$}
\\
}


\begin{abstract}
\noindent The role of weakly bound neutral clusters, such as dineutrons and tetraneutrons, in matter of
high density and high temperature is discussed. Under such conditions, which are characteristic of core--collapse
supernovae, the lifetime of multineutrons may prove to be sufficiently long for them to have a
pronounced effect on the formation of the chemical composition. The influence of the multineutron binding
energy and other nuclear properties on the magnitude of the effect being considered is examined.
\end{abstract}

\maketitle

\section{INTRODUCTION}

A possible involvement of light neutron-rich nuclei
(clusters) in astrophysical processes was considered
by various authors (see, for example, [1, 2]). An unexpectedly
high concentration of light clusters, such as
\isn{H}{4} and \isn{He}{8}, in the central part of collapsing stars was
recently revealed in [3]. These results were recently
confirmed in [4]. In the present study, attention is
given primarily to the possible role of dineutrons and
tetraneutrons under conditions of high temperature
and a high density, as well as under the conditions of a
significant neutronization of matter, these conditions
being characteristic of supernovae.

The dineutron and tetraneutron are possible quasi-bound
states of several neutrons [5]. A short-term
weakly bound state (of binding energy about
70~keV) of two neutrons (dineutron) may arise owing
to the interaction of neutron magnetic moments, for
example, in the (\isn{\it T}{}, \isn{\it p}{}) reaction where the participant
triton transfers two its neutrons to the target nucleus
or in other reactions involving tritium or deuterium:
(\isn{\it T}{}, \isn{\it T}{}),
(\isn{\it D}{}, \isn{\it T}{}), (\isn{\it D}{}, \isn{\it n}{}). The dineutron lifetime
is about a nuclear lifetime; in all probability [6], the
dineutron may exist as a resonance or as a weakly
bound system within the range of nuclear forces: in
the neutron halo of highly neutron-rich nuclei [6-10]
or in the neutron-star crust. In some studies, it is
indicated that, in the beta decay of highly neutron-rich
nuclei, where the emission of several neutrons is
energetically favorable, two neutrons may be emitted
not only sequentially but also in the form of a dineutron
cluster [11]. Two neutrons may also be bound in
a neutron halo of light nuclei [7].

For the first time, experiment devoted to searches
for dineutrons were performed in 1948 [12, 13]. The
first estimations of the dineutron binding energy were
performed at the same time, $\WI{Q}{ 2n} \sim 0.7\pm 0.2$~MeV,
and the dineutron lifetime with respect to beta decay
was also assessed. Experiments aimed at dineutron
searches were performed with the aid of various reactions
involving the fusion of light nuclei [14-18] and
the decay of heavy nuclei [19]. For example, Spyrou
and his coauthors [20] reported on the detection of
a dineutron in the decay of a \isn{Be}{16} nucleus (in its
decay, the emission of one neutron is impossible) on
the basis of recording two delayed neutrons emitted at
a small angle.

In addition to dineutrons and tetraneutrons (see [5,
21-23] and references therein), other possible weakly
bound multineutron states are also discussed in
the literature [19, 23-25]. There is presently no
unambiguous answer to the question of whether a
tetraneutron exists in the form of a resonance or
a bound state. Nowadays, experiments are being
performed with the aim of observing a tetraneutron
in predominantly three reactions: (i) the induced
fission of \isn{U}{238} [26], (ii) the breakup of \isn{Be}{14} to
\isn{Be}{10} and \isn{\it n}{4} [27] (this experiment was confirmed
by calculations reported in [28, 29]), and (iii) the
reaction \isn{He}{4}(\isn{He}{8}, \isn{Be}{8})\isn{\it n}{4} [30]. However, theoretical
calculations on the basis of modern models of two- and
three-nucleon interaction do not provide an
unambiguous proof of the existence of the tetraneutron
[28]. If the experimental results obtained
in [27, 30] are confirmed together with the arguments
presented in [19, 31] in support of the existence of
neutral clusters containing not less than six neutrons,
this would require revising modern theoretical models
of nuclear forces [32].

The foregoing brings about the question of when
and where these weakly bound states may manifest
themselves. Under various conditions, reactions involving
a dineutron (as well as other neutral clusters)
may have a substantial effect on the results of nucleosynthesis
(from primordial to equilibrium nucleosynthesis),
changing the yield of product nuclides:
\isn{\it n}{2}(p, {\it n}){\it D}, {\it D}(\isn{\it n}{2}, {\it n}){\it T},
\isn{He}{3}(\isn{\it n}{2}, {\it n})\isn{He}{4},
   \isn{He}{3}(\isn{\it n}{2},{\it D}){\it T} и \isn{Be}{7}(\isn{\it n}{2},{\it n})\isn{He}{4}. From an analysis of reactions
with a dineutron in primordial nucleosynthesis, it was
found [2] that, even if variations in fundamental constants
(for example, in the pion mass) led to a change
in the dineutron binding energy, this energy did not
exceed 2.5~MeV within the first minutes after the Bing
Bang. Otherwise, the observed abundances of helium
and deuterium in the Universe would be different.

In the present article, we will describe the equation
of state for matter (Section 2) and consider the role
of neutron clusters in the formation of the chemical
composition of dense and hot supernova stellar matter
(Section 3). We will also discuss those parameters
on which the possible effect of a multineutron may
be especially strong (Section 4). In addition, we
will determine the dependence of the results on the
binding energy of the multineutrons being considered
(Section 5).

\section{EQUATION OF STATE FOR MATTER}

In order to calculate the properties and chemical
composition of supernova matter, we will use two
equations of state that were described in detail in [33,
34]. Both are based on the use of the approximation
of nuclear statistical equilibrium, which is valid
at temperatures that satisfy the condition $T\gtrsim 3\times 10^{9}~\mbox{K}$. Under these conditions, all direct and inverse
nuclear reactions are in equilibrium, in which case the
chemical potential of an arbitrary nucleus with mass
number $A$ and charge number $Z$ is straightforwardly
expressed in terms of the neutron, $\WI{\mu}{ n}$, and proton, $\WI{\mu}{ p}$,
chemical potentials as
\begin{equation}
\WI{\mu}{ A,Z}=(A{-}Z)\WI{\mu}{ n}+Z\WI{\mu}{ p}. \label{mu_NSE}
\end{equation}
In the equation of state proposed by Nadyozhin and
Yudin [33] (EoS NY in the following), nuclei are
treated as a perfect Boltzmann gas, while neutrons
and protons are assumed to have a form of a Fermi
gas whose degree of degeneracy is arbitrary. For nuclei,
one takes into account explicitly low-lying levels
known from experiments and highly excited states
through the Fermi gas model. In our calculations, we
include about 400 nuclei --- predominantly the nuclei
of the iron peak and nuclei of light elements.

The equation of state proposed by Blinnikov,
Panov, Rudzsky and Sumiyoshi [34] (EoS BPRS
in the following) takes additionally into account
matter nonideality effects: Coulomb interaction and
interaction of nuclei with surrounding free nucleons.
The latter is especially important at high densities.
The partition functions for nuclei (without allowance
for excited levels) are calculated according to [35]. In
all, about 4500 nuclei were explicitly included in our
calculation.

It should specially be noted that highly neutron-rich
hydrogen and helium isotopes, which, as was
found in [3, 4], may be abundant at high densities
and temperature characteristic of core-collapse supernovae,
were not included in the calculations in
the aforementioned equations of state. This was
done deliberately in order to demonstrate the effect of
multineutrons against ``standard'' sets of nuclei.

In order to find the equilibrium chemical composition,
it is necessary to solve the following set of
equations for the component concentrations $n_i$:
\begin{equation}
\begin{cases}
\sum\limits_i n_i A_i=\WI{n}{ b},\\
\sum\limits_i n_i Z_i=\WI{n}{ e}.
\end{cases}
\label{system}
\end{equation}
Here, summation is performed over all nuclei (including
free nucleons), while $A_i$ and $Z_i$ are, respectively,
the mass and charge numbers of specific nuclei. Further,
$\WI{n}{ b}\equiv\rho/\WI{m}{ u}$ is the baryon concentration ($\rho$ is the
matter density, $\WI{m}{ u}$ is an atomic mass unit). Thus,
the first equation in the set of Eqs. (2) is the condition
of baryon-number conservation. In the second equation
in the set of Eqs. (2), $\WI{n}{ e}$ is the electron concentration;
therefore, the second equation is the condition of
electric neutrality of matter. It is convenient to define
a dimensionless electron concentration (fraction) $\WI{Y}{ e}$
as $\WI{Y}{ e}\equiv
\WI{n}{ e}/\WI{n}{ b}$. In matter featuring equal numbers
of neutrons and protons (helium, \isn{He}{4}; carbon, \isn{C}{12}
etc.),  $\WI{Y}{ e}=\frac{1}{2}$; for iron, \isn{Fe}{56}, we have $\WI{Y}{ e}=\frac{26}{56}$. We
now fix the values of the temperature $T$, density $\rho$,
and electron fraction $\WI{Y}{ e}$. We can then solve the set
of Eqs. (2) with respect to two variables, $\WI{\mu}{ n}$ and $\WI{\mu}{ p}$.
Substituting them into Eqs. (1), we find the chemical
potentials and, hence, the concentrations of all other
components. This solves the problem of determining
the chemical composition of matter. Thus, the equation
of state under conditions of nuclear statistical
equilibrium is fully determined by specifying three
parameters: $\{T,\rho,\WI{Y}{e}\}$.

\section{``STANDARD'' CALCULATION
OF CONCENTRATIONS}

An example of the results obtained from calculations
according to EoS NY with a standard set of
nuclei is given in Fig.~1. For the purposes of illustration,
we chose the moment of time at which the
matter density at the center of the collapsing stellar
core reached a value of about $\rho\approx 3\times
10^{13}$~g~cm$^{-3}$,
in which case a time of about 1 ms remains before
an moment of time of collapse termination (bounce
effect). We choose this moment of time of the collapse
process and the respective profiles of the distribution
of thermodynamic parameters in the stellar core (for
details of the calculation, see [36]) as a characteristic
example that makes it possible to examine all special
features of the distribution of the chemical composition
of matter. At high densities, a substantial
deviation from the ideality of matter arises because of
a strong nuclear interaction between its components,
and the equations of state that we consider become
inaccurate. At a density of $\rho\simeq
10^{14}$~g~cm$^{-3}$, nuclei
disappear via a phase transition to uniform nuclear
matter. Concurrently, the stiffness of the equation of
state increases substantially, the rate of the collapse
process at the center of the star being considered
sharply becomes lower, and there arises a diverging
shock wave leading eventually to the ejection of the
stellar envelope --- that is, to a supernova explosion.

\begin{figure}
\setcaptionmargin{5mm} \onelinecaptionstrue
\center
\includegraphics[width=1\textwidth]{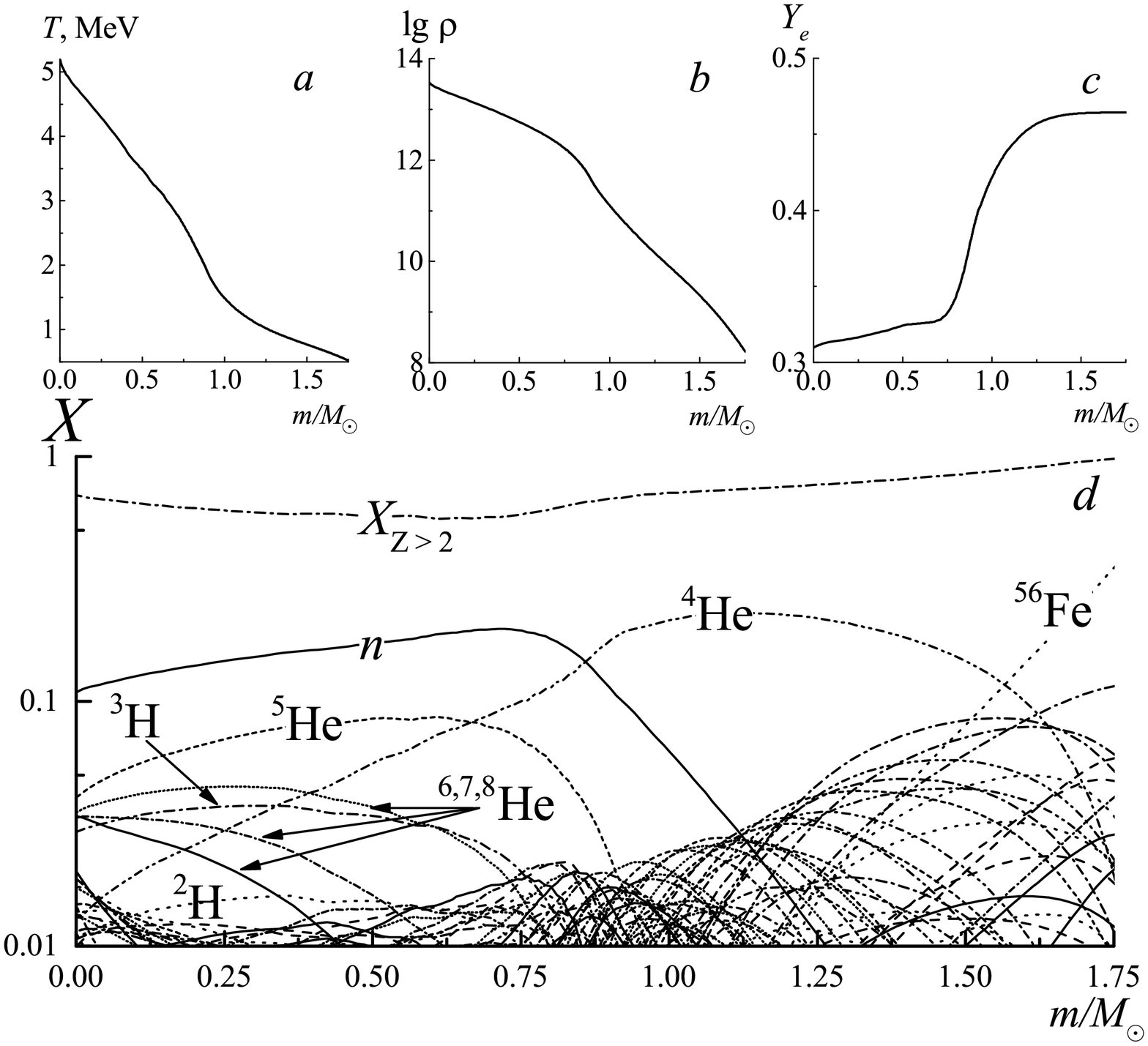}
\vspace{-1cm}
\captionstyle{normal} \caption{Dependencies of the (a) temperature $T$ , (b) logarithm of the density $\rho$, and (c) electron fraction $\WI{Y}{e}$ on the mass coordinate
$m$ in matter of a core-collapse supernova to a time of $t \approx 1$~ms before the bounce. (d) Weight fractions of isotopes $X_i$ (the
isotope species are indicated on the curves) in the case of EoS NY for a standard set of nuclei versus the mass coordinate. }\label{Pix:before_bounce}
\end{figure}
Figures 1a, 1b, and 1c show the distributions
of, respectively, the temperature $T$ (in MeV units:
1 MeV approximately corresponds to $11.6\times 10^{9}$~K);
logarithm of the density (in g~cm$^{-3}$ units), $\lg\rho$; and
the electron fraction  $\WI{Y}{ e}$ in matter versus the mass
coordinate $m$ ($M_\odot$ is the Sun's mass, and $m = 0$ corresponds
to the center of the star). Figure 1d shows
the chemical composition of matter ($X$ is the weight
fraction of an element) at the same moment of time. In
just the same way as for electrons, the dimensionless
concentrations are defined as the ratio $Y_i\equiv n_i/\WI{n}{ b}$,
where $n_i$ is the concentration of the i-th element. For
an element of mass number $A_i$, the weight fraction
$X_i$ is naturally related to its concentration by the
equation $X_i = Y_i A_i$. The dash-dotted line marked by
the symbol $X_{Z>2}$ represents the total weight fraction
of all nuclei for which $Z_i > 2$. As follows from the first
equation in the set of Eqs. (2), the weight fractions $X_i$
satisfy the normalization condition $\sum_i X_i = 1$, where
the sum is taken over all nuclei and free nucleons.

At large values of the coordinate $m$, one can see
remnants of the original iron core of the star. The
growth of the density and temperature as we move
to the center leads to the dissociation of iron-peak
nuclei to ever lighter elements, nucleons, and alpha
particles. In addition, matter simultaneously undergoes
a substantial neutronization (this corresponds
to a decrease in  $\WI{Y}{e}$ to a value of about 0.3 at the
center) accompanied by the appearance of neutron-rich
isotopes of chemical elements and an increase in
their amount. The concentration of free neutrons also
grows. It is noteworthy that, although isotopes of
light elements have large \emph{individual} concentrations
(weight fractions) at the central part of the core, the
total weight fraction of heavy elements (dash-dotted
line) prevails there as well.

\section{CALCULATION WITH ALLOWANCE
FOR MULTINEUTRONS}

Postponing, for the time being (see Section 6 below),
the discussion on the validity of including unbound
states of negative binding energy, dineutrons
and tetraneutrons, in the calculation of the equation
of state in the approximation of nuclear statistical
equilibrium, we will take into account these
neutral clusters in calculating the equation of state
and examine the results obtained in this way. We
will use zero value for the ground-state spin of either
cluster and set the dineutron binding energy to
$\WI{Q}{ 2n}=-66$~keV and the tetraneutron binding energy
to $\WI{Q}{ 4n}=-0.8$~MeV [29].

\begin{figure}
\setcaptionmargin{5mm} \onelinecaptionstrue
\includegraphics[width=\textwidth]{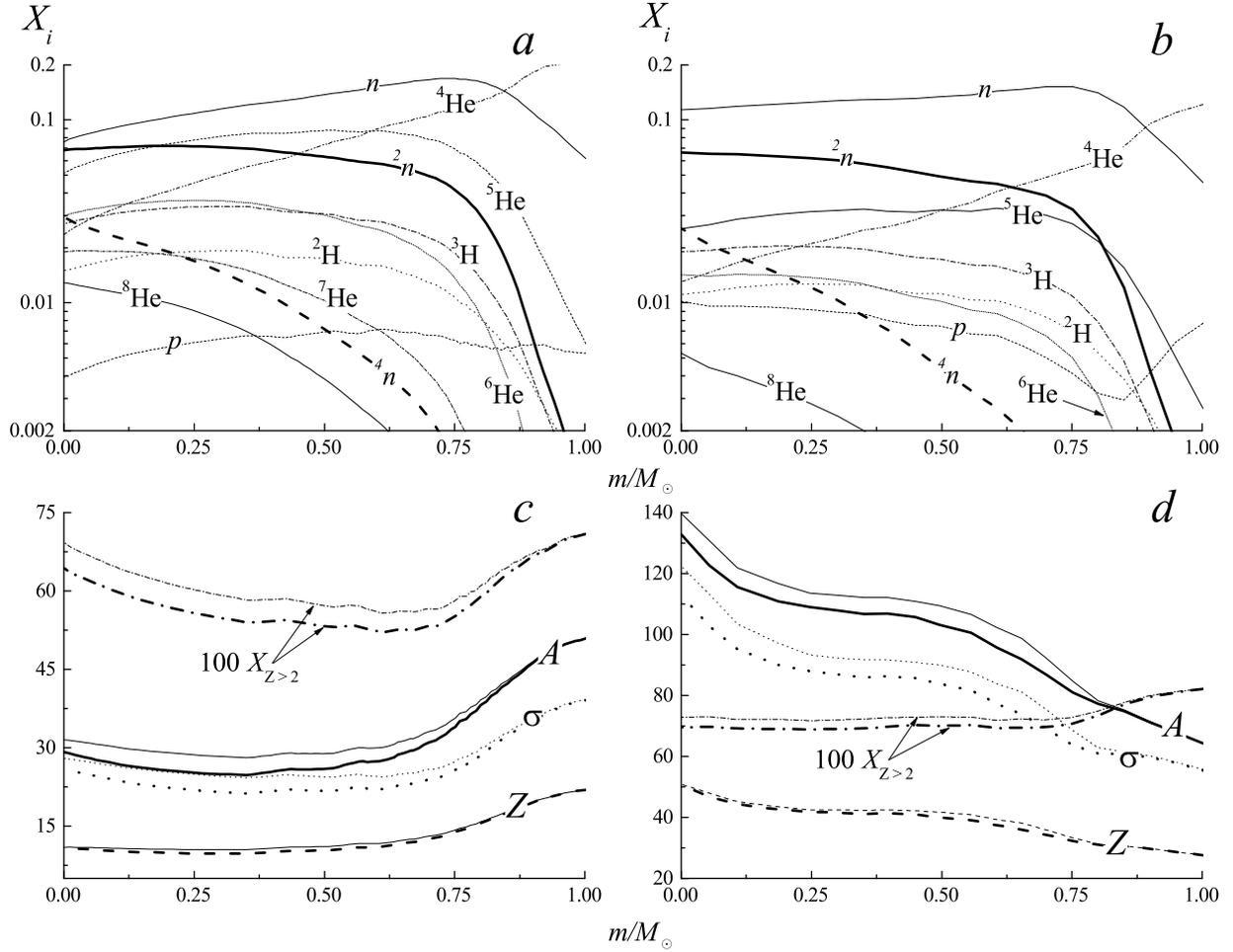}
\vspace{-1cm}
\captionstyle{normal} \caption{(a, b) Weight fractions $X_i$ of $Z \leq 2$ light elements and multineutrons (thick curves) versus the mass coordinate m
in the central region of the collapsing core at the same moment of time as in Fig. 1. The isotope symbols play of the role of
notation for the curves. (c, d) Various integrated EoS features, including the average values of the mass, $A$, and charge, $Z$,
numbers; relative cross section for coherent neutrino scattering on nuclei, $\sigma$ (for more details, see main body of the text); and
weight fraction of $Z > 2$ elements [both without \isn{n}{2} and \isn{n}{4} (thin curves) and with allowance for them (thick curves)]. The
results on display were obtained on the basis of (a, c) EoS NY and (b, d) EoS BPRS. } \label{Pix:with_multin}
\end{figure}
Figure 2 illustrates a comparison of the results
obtained for the equations of state being considered.
Figures 2a and 2c refer to EoS NY, while Figs. 2b
and 2d refer to EoS BPRS. The moment of time and
conditions are identical to those in Fig. 1. Figures 2a
and 2b show the distributions of weight fractions of
$Z_i \leq 2$ light components of matter (in particular, the
thick solid and dashed curves represent the contributions
for dineutrons and tetraneutrons, respectively)
in a central part of the core where $0\leq m/M_\odot\leq 1
$. In
the outer part, the inclusion of multineutrons does not
lead to any significant changes. One can see that, in
the region of $m/M_\odot\lesssim 1$ (which, according to Fig. 1,
corresponds to densities satisfying the condition $\rho\gtrsim
10^{11}$~g~cm$^{-3}$), dineutrons are at least as abundant as
other light nuclei; in the central part of the stellar core,
they are even second to only free neutrons. As for
tetraneutrons, they seem less significant, even though
their contribution grows fast with increasing density.
In principle, the two equations of state used, which
are quite different, lead to a consistent picture in this
respect, despite a difference in details.

We will now proceed to consider Figs. 2c and 2d.
In them, the behavior of several summed EoS features
calculated for the same region of the stellar core with
a standard set of nuclei (thin curves) is compared
with their counterpart for the extended set including
multineutrons (thick curves). We recall that the number
of nuclear species included in the calculations is
strongly different for the two equations of state under
consideration. We begin by examining a quantity with
which we have already dealt in Fig. 1: the dash-dotted
curve represents the total weight fraction of all $Z_i > 2$
``heavy'' nuclei that was multiplied by a factor of 100
for convenience of a presentation. One can see that
the inclusion of multineutrons leads to some decrease
(of about 7\% for EoS NY and about 5\% for EoS
BPRS) in the fraction of heavy nuclei. The symbols
$A$ and $Z$ (solid and dashed curves) stand for their
average mass number and average charge number,
respectively. The averaging in question is performed
according to the rule (see, for example, [37])
\begin{equation}
\langle A\rangle\equiv\frac{\sum_{Z>2}A_i n_i}{\sum_{Z>2}n_i},\quad
\langle Z\rangle\equiv\frac{\sum_{Z>2}Z_i n_i}{\sum_{Z>2}n_i}.
\label{AverageAZ}
\end{equation}

From these figures, it is also clear that the effect of
multineutrons leads to some decrease (of about 5\%)
in the average charge and mass of heavy nuclei. Two
points are in order here. First, the inclusion of multineutrons
affects these quantities indirectly, since, in
the two cases being considered (that is, the calculations
with and without \isn{n}{2} and \isn{n}{4}), averaging is
performed over the same set of nuclei (those for which
$Z_i > 2$). Second, a significant difference in the values
of $\langle A\rangle$ and $\langle Z\rangle$ for the two equations of state being
considered is noteworthy. For EoS NY, the average
charge number in the central region is slightly greater
than 10, while the average mass number is about
30, but, for Eos BPRS, $\langle Z\rangle\sim 40$ while  $\langle A
\rangle\gtrsim 100$.
This is because EoS NY takes into account a very
restricted region of nuclei that covers only $Z < 36$ and
$A < 83$ nuclei. It follows that, here, the equilibrium
chemical composition at high densities is a mixture of
the iron-peak nuclei and nuclei of light neutron-rich
elements. Not only is the region of nuclei considered
in EoS BPRS wider by a factor of ten, but it also
includes heavy neutron-rich nuclei, which prevail at
high densities, leading to substantially greater values
of  $\langle A \rangle$ and $\langle Z\rangle$. Here, it is of importance that the
above two equations of state, which correspond to so
different a basic chemical composition and to different
underlying microscopic physics, make qualitatively
similar predictions for the behavior of multineutrons
in the region being considered.

The expression
\begin{equation}
\sigma\equiv\sum\limits_{\mathrm{all}}Y_i
A_i^2=\sum\limits_{\mathrm{all}}X_i A_i, \label{sigma}
\end{equation}
where summation is performed over all nuclei; free
nucleons; and multineutrons, if any, is the last
summed quantity whose behavior is depicted in Fig. 2
(dotted curve marked by the symbol $\sigma$). Particular
attention to this quantity is motivated by the following
argument: neutrinos play a dominant role in energy
transfer during the process of collapse of the stellar
core. It is intense flows of neutrinos of all flavors that
carry away the overwhelming portion (about 99\%) of
the whole deposited energy. Therefore, the energy
fraction associated with the ejection of the stellar
envelope and with the photon flux, which is precisely
what we observe as a supernova explosion is less than
one percent. It turns out that, under the conditions
being considered, coherent neutrino scattering on
nuclei is one of the dominant processes of neutrino
interaction with matter. Its cross section, $\WI{\sigma}{cs}$, is
approximately in direct proportion to the square of the
mass number of the nucleus involved: $\WI{\sigma}{cs}(A)\propto A^2$.
It is the averaging of precisely this cross section
over the chemical composition of matter that leads
to the quantity $\sigma$ in Eq. (4). In contrast to the
aforementioned average mass, $\langle A \rangle$, and charge, $\langle Z\rangle$,
numbers of heavy nuclei, the parameter $\sigma$ additionally
includes a direct contribution of multineutrons. One
can see that a relative decrease in $\sigma$ is also moderately
small, but, somewhere, it reaches 10\%.

Summarizing the results obtained in this section,
we can say that the \isn{\it n}{2} and \isn{\it n}{4} effect on summed EoS
features is relatively small, but, possibly, a consistent
inclusion of other super-heavy isotopes of light elements
as well would lead to some noticeable changes
in processes of neutrino production and propagation,
if not in collapse dynamics, and may affect the chemical
composition in the period after the bounce of the
core.

\section{SENSITIVITY TO PARAMETERS}

\begin{figure}
\setcaptionmargin{5mm} \onelinecaptionstrue
\includegraphics[width=1\textwidth]{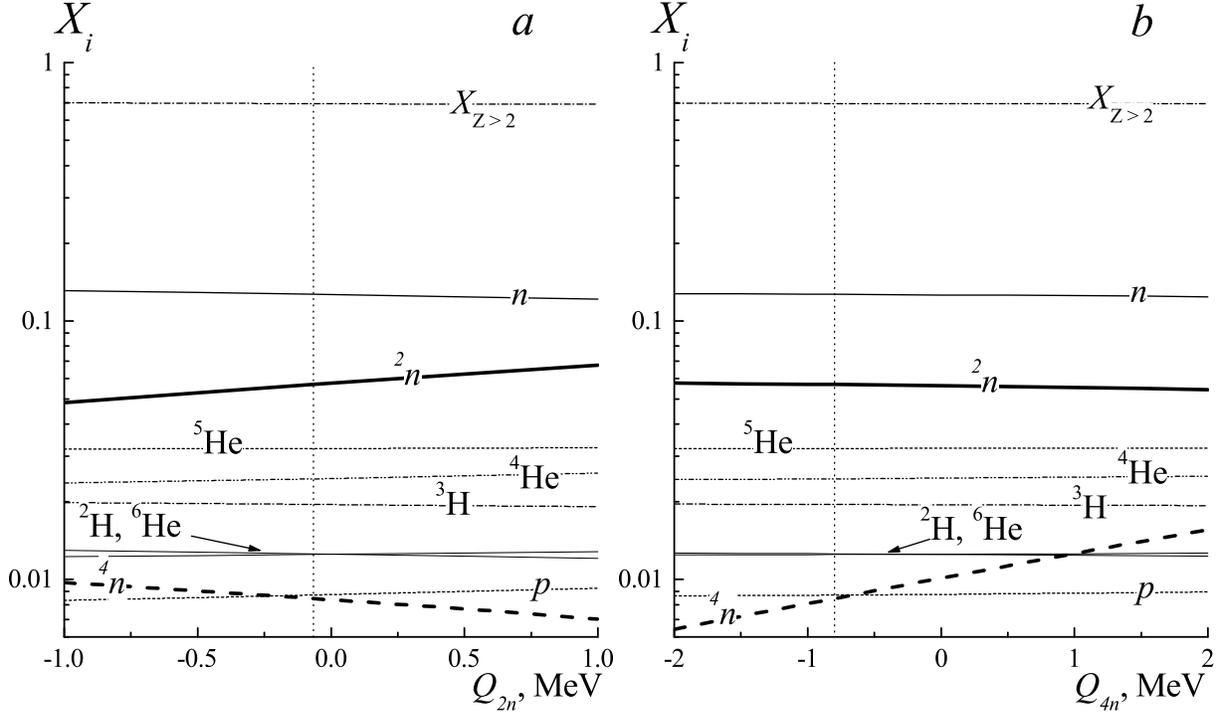}
\vspace{-1cm}
\captionstyle{normal} \caption{Weight fractions of $Z \leq 2$ light nuclei and total weight fraction $X_{Z>2}$ of heavy nuclei versus the (a) dineutron
binding energy $\WI{Q}{2n}$ and (b) tetraneutron binding energy $\WI{Q}{4n}$. The respective calculation was based on EoS BPRS, and
the thermodynamic parameters were set to $T = 4$~MeV, $\rho = 10^{13}$~g~cm$^{-3}$, and $\WI{Y}{e} = 0.32$ } \label{Pix:VarQ}
\end{figure}
Since exact values of multineutron parameters
are unknown, it is of importance to study the sensitivity
of our results to variations in their specific
values --- first of all, to variations in the binding energy.
In order to do this, we have performed the
following calculation on the basis of EoS BPRS.
For the specific thermodynamic-parameter values of
$T = 4$~MeV, $\rho = 10^{13}$~g~cm$^{-3}$, and $\WI{Y}{e} = 0.32$ (this
approximately corresponds to the conditions in Fig. 1
at $m \approx 0.34 M_\odot$), we have calculated the chemical
composition of matter at various values used for the
\isn{\it n}{2} and \isn{\it n}{4} binding energies. The results are given
in Fig. 3. The weight fractions of $Z \leq 2$ light nuclei
and the total weight fraction of heavy nuclei, $X_{Z>2}$,
are shown there versus the dineutron binding energy
$\WI{Q}{2n}$ (Fig. 3a) and the tetraneutron binding energy
$\WI{Q}{4n}$ (Fig. 3b). A negative binding energy naturally
corresponds to an unbound state. We note, in passing
that the experiment reported in [10] revealed a
binding energy for light nuclei in excess of theoretical
predictions. We varied the \isn{\it n}{2} binding energy over
the range of $-1\leq\WI{Q}{ 2n}\leq 1$~MeV and the \isn{\it n}{4} binding
energy over the range of $-2\leq\WI{Q}{ 4n}\leq 2$~MeV. The
``commonly accepted'' values of $\WI{Q}{ 2n}=-0.066$~MeV
and $\WI{Q}{ 4n}=-0.8$~MeV are indicated by vertical dotted
lines. The effect is seen to be quite moderate; it is the
most pronounced for tetraneutrons. First of all, variations
in the binding energies affect the multineutrons
themselves, the concentrations of the other components
remaining virtually unchanged. Thus, we can
draw the conclusion that the calculated multineutron
concentration is weakly sensitive to variations in the
binding energies at preset values of the temperature
and density. This comes as no surprise since only
at high values of the density and temperature ($T =
4$~MeV in the example being considered) do multineutrons
appear in matter. All thermodynamic quantities
depend only on the ratio of the binding energy to
temperature, $q\equiv Q/T$; in the region being considered,
$|q|<1$ is a small parameter, taking values in the
vicinity of zero.

\section{DISCUSSION}
\label{Discussion}

The explosion of a massive star as a supernova
is initiated by the gravitational collapse of its core
that underwent evolution. One possible explosion
mechanism is based on energy transfer from a hot
protoneutron star to a layer above its surface. This
energy released from the core leads to the ejection
of the envelope [38, 39]. The development of the
explosion is accompanied by the deleptonization of
protoneutron-star matter via neutrino emission over
a time scale of about 10 to 30 s [40]. The emission
of the bulk of neutrinos occurs at this stage,
and it is of paramount importance to determine the
spectrum of emitted neutrinos and their luminosity.
An efflux of matter that forms a so-called hot wind
occurs simultaneously during this explosion phase
under the effect of neutrino-induced heating of the
protoneutron--star surface in the deleptonization process
[41]. Conditions for the development of nucleosynthesis
of heavy elements produced under the
effect of neutrons arise in this wind. Therefore, a
correct calculation of neutrino transport is of crucial
importance since this will make it possible to describe
adequately the deleptonization process [42]. For this,
one needs an adequate equation of state for matter
and realistic reaction rates [43]. The importance of
taking into account light clusters, such as deuterium
and tritium, \isn{H}{2,3}; helium, \isn{He}{3,4}; etc., in calculating
the equation of state for matter has been discussed
for a long time (see, for example, [44, 45] and references
therein), but, to the best of our knowledge,
no attention has thus far been given to the ``exotic''
possibility of taking into account multineutrons that
was considered in the present article.

Let us now discuss known experimental data on
multineutrons. The value that we use for the dineutron
binding energy is negative ($\WI{Q}{ 2n}=-0.066$~MeV [5]). The requirement of agreement
between the results of calculations and observations
in simulating primordial nucleosynthesis in the Bing
Bang [2] sets an upper limit of 2.5 MeV on the
dineutron binding energy. Moreover, investigations of
elastic neutron scattering on deuterons, ({\it n}, {\it D}), [46]
give grounds to conclude that available data are
incompatible with the existence of the dineutron
whose binding energy is greater than 100 keV. According
to a large number of studies [28, 30], the
tetraneutron is likely to have a positive binding energy
less than 3.1 MeV [27]; together with the dineutron,
not only can the tetraneutron change the composition
of matter during collapse, leading to a somewhat
different input composition for the subsequent alpha
process or r--process, but it is also able to affect the
transparency of the neutrino-sphere. Therefore, the
ranges that we chose for values of the dineutron and
tetraneutron binding energies (see Fig. 3) reflect their
modern estimates.

In central regions of collapsing stellar cores, matter
is under conditions of nuclear statistical equilibrium,
in which case all direct and inverse reactions
are in equilibrium. In addition, it is assumed that
all possible states, including both bound states and
states that have a resonance character and which lie
in a continuum, should be taken into account in the
calculation. A multineutron lifetime in the range of
($10^{-12}$---$10^{-21}$)~s is likely to be sufficient for examining
the possible involvement of these states in the determination
of the properties of hot and dense astrophysical
nuclear plasma. We will now discuss this point
in more detail. We denote by $\tau$ the lifetime of the
multineutron. Its decay rate is then  $\WI{n}{mn}/\tau$, where $\WI{n}{mn}$
is its equilibrium concentration. A multineutron may
originate, for example, from various collision--induced
reactions; that is, its production rate is $n_i n_j\langle\sigma v\rangle$,
where $n_{i,j}$ stands for the concentrations of colliding
particles, $v$ is their relative velocity, and $\sigma$ is the
cross section for the respective process. There are
many such reactions, and each makes an individual
contribution to multineutron production. For the sake
of simplicity, we disregard similar reactions leading
to multineutron disintegration, thereby obtaining an
upper limit on the abundance of multineutrons. For
characteristic values of thermodynamic parameters
in the region of our interest (see Fig. 1), we take
$T\simeq 5$~MeV and $\rho\simeq 10^{13}$~g~cm$^{-3}$. The velocity is
$v\simeq\sqrt{kT/\WI{m}{ u}}$. From the balance of reactions, we can
then obtain an order-of-magnitude estimate of the
mass fraction of multineutrons in equilibrium. The
result is
\begin{equation}
\WI{X}{mn}\simeq  10^{22}\cdot\tau(\mbox{с})\ \rho_{13}\sqrt{T_5}\sum\limits_{i,j} Y_i Y_j\WI{\sigma}{ b}(i,j),\label{Xmn}
\end{equation}
where $\rho_{13}\equiv\rho\times 10^{-13}$, $T_5\equiv kT/5$~MeV, and the
cross sections $\WI{\sigma}{ b}$ are measured in barns. The sum
in (5) is taken over all multineutron-production channels.
The above estimates of lifetimes are compatible
(depending on cross-section and lifetime values)
with the significant multineutron contribution to the
chemical composition of supernova matter under the
conditions being considered.

We have shown that plasmas of supernova matter
at high densities and temperature and under conditions
of strong neutronization of matter may receive
a significant contribution not only from neutron-rich
hydrogen and helium isotopes [3, 4] (which we have
not included in the present calculation) but also from
light purely neutron clusters, the more so as multineutron
systems containing six or more neutrons may
be bound [10, 19]. Here, it is important that, both
for weakly bound states and for quasistable states
possibly arising at the protoneutron-star surface, the
abundances of light neutron clusters change insignificantly
in response to a change of a few hundred keV
units in the binding energy. In conclusion, we would
like to emphasize that the present calculations are
only indicative of the feasibility and potential importance
of ``exotic'' multineutron states in the calculation
of the equation of state for supernovae. The
problem of their real contribution should be solved on
the basis of new theoretical and experimental data on
respective lifetimes and production cross sections [see
Eq. (5)].

\section{ACKNOWLEDGMENTS}

We are grateful to D.K. Nadyozhin and S.I. Blinnikov
for their participation in discussions of the results
of this study and their interest in it and to
A.G. Doroshkevich for enlightening comments.

This work was supported by Russian Foundation
for Basic Research (project no. 18-29-21019mk).

\centerline{ REFERENCES}


%
%

\end{document}